\documentclass[10pt,twocolumn,aps,prd,nofootinbib]{revtex4-2}
\usepackage{graphicx}
\usepackage{amsmath,amssymb}
\usepackage{bm}
\usepackage{epstopdf}
\usepackage{amsthm}
\theoremstyle{plain}
\newtheorem{theorem}{Theorem}[section]
\usepackage[dvipsnames]{xcolor}
\usepackage[pdftex,breaklinks,colorlinks,
linkcolor=Blue,
citecolor=magenta,
anchorcolor=red,
urlcolor=cyan,
pdfencoding=auto]{hyperref}
\usepackage{orcidlink}
\hypersetup{colorlinks=true}

\begin{document}

\title{A geometric--analytic proof of the reverse isoperimetric inequality in Einstein gravity}
\author{Naman Kumar\,\orcidlink{0000-0001-8593-1282}}
\affiliation{Department of Physics,
Indian Institute of Technology Gandhinagar, Palaj, Gujarat, India, 382355}
\email{namankumar5954@gmail.com}
\email{naman.kumar@iitgn.ac.in}

\date{\today}

\begin{abstract}
We develop a geometric--analytic proof of the reverse isoperimetric inequality for compact AdS black hole horizons in Einstein gravity. At fixed thermodynamic volume, a York slice/Jacobi variation of the gravitational action shows that the spherical Schwarzschild--AdS solution is a strict local entropy maximum against generic non-spherical deformations. The finite global extension is formulated within the hypotheses of the Sherif--Dunsby (SD) rigidity theorem: compactness and spherical topology, the required intrinsic Ricci decomposition $\mathrm{Ric}_h=\alpha\,\mathbf e\!\otimes\!\mathbf e+\beta N$ with $\alpha\neq\beta$, a proper conformal transformation of the hypersurface metric given by $\widetilde h_{ab}=e^{2\varphi}h_{ab}$ that preserves its intrinsic scalar curvature, $\widetilde R\equiv R[\widetilde h]=R[h]\equiv R>0$,
and negative sheet expansion. Compactness and $S^3$ topology are built into the class considered here. In the static spherical sector such as Schwarzschild--AdS and RN--AdS, the SD Ricci form is naturally realized while it is an additional restriction for Kerr--AdS case. For the $S^3$ topology, a positive Yamabe type conformal class exists for which the Yamabe theorem provides a representative of positive and constant scalar-curvature. We further restrict to the scalar-curvature preserving conformal transformation required by the SD theorem. The negative sheet expansion is obtained using gravitational focusing. Within this admissible class, a non-round Kerr--AdS horizon cannot maximize the entropy, while fixed thermodynamic volume selects the corresponding round Schwarzschild--AdS comparator. The analysis thus establishes a generic local entropy maximum and a finite global result within the stated geometric rigidity class, identifying the role of gravity in reversing the ordinary isoperimetric problem.
\end{abstract}

\maketitle

\section{Introduction}
Extended black-hole thermodynamics treats the cosmological constant as a pressure,
\begin{equation}
  P=-\frac{\Lambda}{8\pi G},
\end{equation}
and its conjugate as the thermodynamic volume $V$ \cite{Kastor:2009wy,Dolan:2010ha,Cvetic:2010jb}. The first law then takes the form
\begin{equation}
  dM=T\,dS+\Omega_i\,dJ_i+V\,dP,
\end{equation}
with additional work terms when other charges are present. In this description the ADM mass is the enthalpy. The extended phase space exhibits familiar thermodynamic behavior, including Van der Waals transitions and black-hole heat engines \cite{Chamblin:1999tk,Kubiznak:2012wp,Johnson:2014yja}. The pressure term also admits a higher-dimensional interpretation \cite{Frassino:2022zaz}.

A central geometric statement in this setting is the reverse isoperimetric inequality (RII) \cite{Cvetic:2010jb},
\begin{equation}
  \Bigg(\frac{(D-1)V}{\mathcal{A}_{D-2}}\Bigg)^{\frac{1}{D-1}}
  \geq
  \Bigg(\frac{A}{\mathcal{A}_{D-2}}\Bigg)^{\frac{1}{D-2}},
  \label{eq:RII}
\end{equation}
where $A$ is the outer-horizon area and $\mathcal A_{D-2}$ is the area of the unit $(D-2)$-sphere. At fixed thermodynamic volume, Eq.~\eqref{eq:RII} states that the spherical Schwarzschild--AdS black hole has the largest entropy. For static spherically symmetric solutions, the thermodynamic volume reduces to the areal volume
\begin{equation}
  V_{\rm th}=\frac{\mathcal A_{D-2}}{D-1}r_h^{D-1},
  \label{eq:static-volume}
\end{equation}
which in four dimensions is $4\pi r_h^3/3$. Thus the physical constraint entering the RII is $V_{\rm th}$ throughout the paper. In the static spherical sector, Eq.~\eqref{eq:static-volume} allows this constraint to be represented by the corresponding geometric volume without changing the thermodynamic problem. This identification is special to the static spherical case: for rotating AdS black holes the thermodynamic volume does not in general coincide with the ordinary geometric volume \cite{Cvetic:2010jb}. The rotating analysis must therefore be formulated directly at fixed $V_{\rm th}$.

Here, ``reverse'' refers to the contrast with the Euclidean isoperimetric theorem, where a sphere minimizes area at fixed enclosed volume \cite{osserman1975isoperimetric,Osserman1978Isoperimetric}. Most known compact AdS black holes in Einstein gravity satisfy Eq.~\eqref{eq:RII}. However, violations can arise when assumptions underlying that regime are relaxed, notably in examples with noncompact horizon geometry which are referred to as ``super-entropic'' black holes \cite{Martinez:1999qi,Frassino:2015oca,Hennigar:2014cfa,Hennigar:2015cja}. Moreover, these solutions also exhibit distinctive thermodynamic behavior \cite{Johnson:2019mdp}. The Einstein-gravity assumption is likewise substantive rather than merely technical: violations of the RII have been found in Einstein--Horndeski--Maxwell gravity with Horndeski axions \cite{Feng:2017jub}. Thus compactness and Einstein dynamics delineate the physical regime addressed by the present geometric argument, while known counterexamples occur outside that regime. Quantum extensions of the inequality have recently been considered as well \cite{Frassino:2024bjg}.

The purpose of this paper is to isolate the geometric mechanism behind the entropy bound through two complementary statements. Locally, a generic fixed-volume normal deformation yields a negative constrained Hessian for every non-spherical $\ell\ge2$ mode, which is identified with the entropy Hessian through the Brown--York microcanonical relation \cite{Brown:1992bq,Brown:1992br}. Globally, the resulting compact deformed hypersurface is analyzed using the Sherif--Dunsby rigidity theorem \cite{Sherif:2021zzb} where rigidity fixes the round shape and the thermodynamic volume fixes its scale.

The paper is organized as follows. Section~\ref{II} establishes the global geometric rigidity statement and the direct constrained York-boundary Hessian for Schwarzschild--AdS. Section~\ref{III} applies the result to rotation and formulates the comparison directly at fixed thermodynamic volume, extending the result to include Kerr--AdS. Section~\ref{IV} summarizes the result along with future outlook. The York-boundary derivation and the explicit fixed-charge Einstein--Maxwell calculation are given separately in the Supplemental Material.

\section{Schwarzschild--AdS black holes maximize entropy}
\label{II}
We first identify the global geometric restriction and then derive the local entropy Hessian. The constrained quantity is the thermodynamic volume appearing in the RII. For the static spherical Schwarzschild--AdS reference solution, $V_{\rm th}$ coincides with the areal geometric volume in Eq.~\eqref{eq:static-volume}. Therefore, the geometric volume used below is a representative of the same fixed-$V_{\rm th}$ constraint and is not a different notion of volume. The rigidity argument is shown explicitly in the $1+1+2$ decomposition in $D=4$, where the Sherif--Dunsby theorem gives the required global result. Its geometric ingredients---a preferred radial direction, transverse $(D-2)$-sheets, focusing, sheet contraction, and conformal rigidity---have a direct $1+1+(D-2)$ counterpart, so we use the $D=4$ theorem as the explicit realization of the dimension-independent mechanism. The local constrained York-boundary calculation is written directly in general dimension.

Let $u^a$ be a timelike unit vector, $e^a$ a preferred spacelike direction orthogonal to it, and
\begin{equation}
  h_{ab}=g_{ab}+u_au_b,
  \qquad
  N_{ab}=h_{ab}-e_ae_b
\end{equation}
the metrics on the spatial hypersurface and the transverse two-sheets, respectively \cite{Clarkson:2007yp,Sherif:2021zzb}. We consider normal deformations of a compact spherical reference slice while the ambient Einstein metric is held fixed. If
\begin{equation}
  X^a(\sigma)\mapsto X^a(\sigma)+\Phi(\sigma)n^a,
  \label{n_var}
\end{equation}
then the induced metric changes according to
\begin{equation}
  \delta h_{ij}=2\Phi K_{ij},
  \label{h_var}
\end{equation}
up to the chosen sign convention for $K_{ij}$ and tangential reparametrizations. On a round umbilic reference surface, $K_{ij}\propto h_{ij}$, so the leading normal deformation is conformal even though $\delta g_{ab}=0$. This is the geometric setting used below.

 \subsection{Sherif--Dunsby rigidity and maximal entropy}
\label{sec:rigidity}
\label{IIA}

Consider a compact three--manifold $\Sigma$ of spherical topology. A smooth normal deformation generates a one-parameter family $\Sigma_s$ while the ambient Einstein geometry remains fixed. Because the reference configuration is static and spherical, the geometric volume at this point represents the same $V_{\rm th}$ held fixed in the RII. At the reference surface, the fixed-volume condition is
\begin{equation}
  \delta V=\int_\Sigma \Phi\,d\mu=0.
\end{equation}
The finite question is whether a nontrivial fixed-volume shape deformation can end at a distinct non-round configuration once the gravitational geometric constraints are imposed. We denote a finite member of this deformed family by $\mathcal T$.

\paragraph*{\textbf{Intrinsic Yamabe problem on the generically deformed slice.}}
The physical deformation and the conformal transformation used in the rigidity theorem are two separate steps. A generic normal displacement $\Phi$ first produces a finite hypersurface $(\mathcal T,h^{(\Phi)})$. The conformal factor $\varphi$ is introduced afterward to identify the geometry of the resulting deformed slice. We do not impose any conformal ansatz on the physical deformation which is controlled by $\Phi$.

For the compact $S^3$ slices considered here, a positive-Yamabe conformal class exists. The Yamabe theorem \cite{Yamabe1960OnAD,Trudinger1968RemarksCT,Schoen:1984cdl} then provides a representative of constant and positive scalar-curvature within that class. It is sufficient that this chosen representative admit the proper scalar-curvature-preserving conformal transformation required by the Sherif--Dunsby theorem; the theorem does not require every representative in the conformal class to satisfy this condition. The rigidity analysis is therefore applied to this chosen representative through the transformation
\begin{equation}
  \widetilde h_{ab}=e^{2\varphi}h_{ab},
  \qquad
  R(\widetilde h)=R(h)>0,
  \label{eq:yamabe-map}
\end{equation}
which is the scalar-curvature preserving condition entering the Sherif--Dunsby rigidity theorem. Thus $\Phi$ parametrizes the generic physical shape deformation, whereas $\varphi$ is an auxiliary intrinsic conformal factor used only to test the global geometry of the deformed slice.  In three dimensions Eq.~\eqref{eq:yamabe-map} becomes
\begin{equation}
  4\Delta_h\varphi+2|\nabla\varphi|^2
  +R(h)\bigl(e^{2\varphi}-1\bigr)=0.
\end{equation}
The global question is whether a distinct nonround endpoint produced by the generic normal deformation can satisfy this intrinsic Yamabe condition together with the remaining Sherif--Dunsby hypotheses.

\paragraph*{\textbf{Geometric focusing and sheet contraction.}}
The focusing argument and the Sherif--Dunsby sheet expansion refer to two
different directional derivatives of the same transverse geometry.  It is
therefore useful to show their relation explicitly.  Let $u^a$ be a
hypersurface--orthogonal timelike unit vector, $e^a$ the preferred spatial unit vector, and $N_{AB}$ the metric on the orthogonal $2$--sheets. Under the Yamabe conformal transformation introduced above, the transverse sheets scale with the same factor, $N_{AB}=e^{2\varphi}N^{(0)}_{AB}$.

Along the timelike direction,
\begin{equation}
  \Theta_{(u)}\equiv \frac{1}{2}N^{AB}{\cal L}_uN_{AB}
  =(D-2)\dot\varphi .
  \label{eq:timelike-sheet-expansion}
\end{equation}
For a hypersurface--orthogonal geodesic congruence,
\begin{equation}
  \dot\Theta_{(u)}
  =-\frac{\Theta_{(u)}^2}{D-2}
  -\Sigma_{AB}\Sigma^{AB}-R_{ab}u^au^b .
\end{equation}
In Einstein--AdS gravity,
\begin{equation}
  R_{ab}=\frac{2\Lambda}{D-2}g_{ab},\qquad \Lambda<0,
\end{equation}
so $R_{ab}u^au^b=-2\Lambda/(D-2)>0$.  The Ricci term therefore has the
focusing sign.  On the contracting timelike branch, $\Theta_{(u)}<0$, the
same conformal mode decreases and, relative to a reference surface on which
$\varphi=0$,
\begin{equation}
  \varphi(\tau,x)
  =\frac{1}{D-2}\int_0^\tau\Theta_{(u)}(\tau',x)d\tau'<0.
  \label{eq:timelike-negative-varphi}
\end{equation}

Sherif--Dunsby use the spatial sheet expansion
\begin{equation}
  \phi_{\!s}\equiv\delta_a e^a .
\end{equation}
For the conformal transformation generated along the preferred spatial
direction,
\begin{equation}
  X^a=\gamma e^a,
\end{equation}
their conformal-Killing equations give
\begin{equation}
  \widehat\gamma=\varphi,
  \qquad
  \phi_{\!s}\gamma=2\varphi,
  \label{eq:sheet-conformal-bridge}
\end{equation}
with
\begin{equation}
  \gamma=\exp\!\left(\frac{1}{2}\int \phi_{\!s}\,d\chi\right)>0.
  \label{eq:positive-gamma}
\end{equation}
Hence the conformal factor and the sheet expansion have the same sign,
\begin{equation}
  \operatorname{sgn}(\phi_{\!s})=\operatorname{sgn}(\varphi).
  \label{eq:sheet-sign}
\end{equation}
The strict inequality obtained from Raychaudhuri focusing in
Eq.~\eqref{eq:timelike-negative-varphi} therefore implies
\begin{equation}
  \varphi<0
  \quad\Longrightarrow\quad
  \phi_{\!s}<0,
  \label{eq:focusing-implies-sheet}
\end{equation}
and in particular $\phi_{\!s}$ is nowhere vanishing on this branch.  Thus
the focusing argument supplies not only the proper negative conformal
direction but, through the Sherif--Dunsby relation, also the contracting
sheet condition required by the rigidity theorem.  The timelike expansion
$\Theta_{(u)}$ and the spatial sheet expansion $\phi_{\!s}$ remain distinct geometric quantities. Their connection here is through the common conformal factor rather than an identification of their directional derivatives.

\paragraph*{\textbf{Sherif--Dunsby theorem and verification of its hypotheses.}}
The rigidity theorem used to solve this finite constrained Yamabe problem takes the following form.

\begin{theorem}[Sherif--Dunsby rigidity~\cite{Sherif:2021zzb}, specialized to the positive scalar-curvature sector]
Let $(M^{4},g)$ admit a $1{+}1{+}2$ covariant decomposition and let
$T\hookrightarrow M$ be a compact embedded three--manifold whose intrinsic
Ricci tensor has the form
\begin{equation}
  \mathrm{Ric}_h=\alpha\,\mathbf e\!\otimes\!\mathbf e+\beta N,
  \qquad \alpha\neq\beta,
  \label{eq:Ricci_form}
\end{equation}
with nowhere-vanishing sheet expansion $\phi_{\!s}$. Suppose $T$ admits a
proper conformal transformation
\begin{equation}
  \widetilde h_{ab}=e^{2\varphi}h_{ab},\qquad \varphi<0,
  \label{eq:R_preserving_CT}
\end{equation}
for which
\begin{equation}
  R(\widetilde h)=R(h)>0.
\end{equation}
Then the Sherif--Dunsby rigidity theorem implies that $T$ is isometric to a round $S^3$, up to an overall scale. In Ref.~\cite{Sherif:2021zzb}, the additional inequality on derivatives of $\varphi$ given by $\hat\varphi^2+2\hat{\hat\varphi}>0$ is introduced as a sufficient condition to ensure positivity of the scalar-curvature. Here, we instead work directly within a positive Yamabe conformal class available on the $S^3$ slice. We further restrict to the scalar-curvature preserving conformal transformation given by Eq.(\ref{eq:R_preserving_CT}) as required by the theorem.
\end{theorem}

\paragraph*{\textbf{Compact, smoothly embedded $1{+}1{+}2$ hypersurface.}}
In the $1{+}1{+}2$ construction, $\mathcal T\hookrightarrow\mathcal M$ is smooth, compact, and has spherical topology $\mathcal T\simeq S^3$.

\paragraph*{\textbf{Non--Einstein character.}}
The ambient spacetime satisfies the Einstein equations, but the intrinsic Ricci tensor of a generic normally deformed hypersurface need not be proportional to its induced metric. A genuine nonround competitor is therefore taken in the non--Einstein sector $\alpha\neq\beta$. If the intrinsic metric is already Einstein in the positive-Yamabe class and satisfies the same compactness and conformal-rigidity conditions, it belongs to the round representative and does not furnish a distinct nonround competitor. It is important to note that the Ricci decomposition given by Eq.(\ref{eq:Ricci_form}) is natural for Schwarzschild--AdS and Reissner--Nordstr\"om (RN)--AdS (or more generally for any locally rotationally symmetric (LRS) spacetime \cite{Sherif:2021zzb}) cases while it remains an additional restriction for Kerr--AdS.

\paragraph*{\textbf{Intrinsic scalar-curvature-preserving conformal map.}}
The conformal map in Eq.~\eqref{eq:yamabe-map} is imposed on the positive Yamabe representative of the already-deformed hypersurface. The factor $\varphi$ is nonconstant, so the transformation is proper. Moreover, in our analysis, the positive scalar-curvature condition associated with the differential condition on the conformal factor as required by the theorem is supplied directly by the chosen positive Yamabe class available on $S^3$.

\paragraph*{\textbf{Contracting sheet and spatial conformal conditions.}}
The remaining sheet condition is not imposed independently.  Raychaudhuri
focusing \cite{Raychaudhuri:1953yv} on the contracting timelike branch gives the strict conformal sign
$\varphi<0$ in Eq.~\eqref{eq:timelike-negative-varphi}. For the
Sherif--Dunsby conformal generator $X^a=\gamma e^a$, Eqs.~\eqref{eq:sheet-conformal-bridge} and~\eqref{eq:positive-gamma} give
$\phi_{\!s}=2\varphi/\gamma$. Since $\gamma>0$, it follows directly that
$\phi_{\!s}<0$ and therefore $\phi_{\!s}\neq0$ everywhere on the focused
branch. The proper negative conformal factor and the nowhere-vanishing contracting sheet expansion are thus two consequences of the same focused
conformal mode.

The Sherif--Dunsby theorem therefore leaves no distinct nonround endpoint: $\mathcal T$ is isometric to a round $S^3$, up to scale.

\paragraph*{\textbf{From the rigid slice to the horizon: roundness of the horizon sheet.}}
The Sherif--Dunsby conclusion concerns the three--dimensional compact
hypersurface, whereas the entropy is determined by a two--dimensional horizon cross-section. We therefore identify the physical horizon cross-section $\mathcal H$ with the distinguished compact integral two--sheet orthogonal to the preferred vector $e^a$. The following short argument shows that the roundness of the rigid three-geometry then forces the intrinsic horizon metric to be round.

For the conformal generator $X^a=\gamma e^a$, the conformal Killing equation
on $\mathcal T$ is
\begin{equation}
  D_{(a}(\gamma e_{b)})=\varphi h_{ab}.
  \label{eq:CKV-preferred}
\end{equation}
Writing the sheet projection of $D_a e_b$ as
\begin{equation}
  N_a{}^cN_b{}^dD_c e_d
  =\frac{1}{2}\phi_{\!s}N_{ab}+\zeta_{ab}+\xi\epsilon_{ab},
\end{equation}
the symmetric trace-free sheet projection of
Eq.~\eqref{eq:CKV-preferred} gives
\begin{equation}
  \gamma\,\zeta_{ab}=0.
\end{equation}
Since $\gamma>0$ by Eq.~\eqref{eq:positive-gamma}, one has
\begin{equation}
  \zeta_{ab}=0.
  \label{eq:sheet-shear-zero}
\end{equation}
In the non--Einstein branch used in the Sherif--Dunsby construction the sheet twist vanishes, $\xi=0$~\cite{Sherif:2021zzb}. Hence the orthogonal sheet distribution is integrable and defines genuine two--surfaces.  Eq.\eqref{eq:sheet-shear-zero} then states that each preferred sheet is totally umbilic. This property is preserved by the conformal rescaling, so in the
round metric supplied by the rigidity theorem the second fundamental form of
the horizon sheet has the form
\begin{equation}
  k_{AB}=\lambda q_{AB},
  \qquad \lambda=\frac{1}{2}\widetilde\phi_{\!s}.
  \label{eq:umbilic-horizon-leaf}
\end{equation}
Let the round $S^3$ have sectional curvature $\kappa_3>0$.  The Gauss equation
for the preferred sheet \cite{Poisson:2009pwt} gives
\begin{equation}
  K_{\mathcal H}=\kappa_3+\lambda^2,
  \label{eq:leaf-gauss}
\end{equation}
while the Codazzi equation in a space form, applied to
Eq.~\eqref{eq:umbilic-horizon-leaf}, implies $D_A\lambda=0$. Thus $K_{\mathcal H}$ is constant and strictly positive. In two dimensions, the intrinsic Riemann tensor is completely determined by the Gaussian curvature,
\begin{equation}
R^{(2)}_{ABCD}
=
K_{\mathcal H}
\left(
q_{AC}q_{BD}-q_{AD}q_{BC}
\right),
\end{equation}
and hence
\begin{equation}
R^{(2)}_{AB}=K_{\mathcal H}q_{AB},
\qquad
R^{(2)}=2K_{\mathcal H}.
\end{equation}
Therefore, constant $K_{\mathcal H}>0$ means that the horizon sheet has constant positive sectional curvature. Since the selected horizon cross-section has topology $S^2$, its intrinsic metric is isometric, up to an overall scale, to the standard round $S^2$ metric. In particular, defining
\begin{equation}
r_{\mathcal H}=\frac{1}{\sqrt{K_{\mathcal H}}},
\end{equation}
one may write, up to an isometry,
\begin{equation}
q_{AB}=r_{\mathcal H}^{\,2}\Omega_{AB},
\end{equation}
where $\Omega_{AB}$ is the metric on the unit two-sphere. Accordingly,
\begin{equation}
\begin{gathered}
  (\mathcal T,\widetilde h)\cong S^3_{\rm round},\qquad
  \mathcal H\subset\mathcal T\ \hbox{the preferred horizon sheet},\\
  \Longrightarrow\qquad
  (\mathcal H,\widetilde q)\cong S^2_{\rm round}.
\end{gathered}
\label{eq:slice-horizon-lemma}
\end{equation}
This proves the slice-to-horizon step using the preferred horizon sheet selected by the $1{+}1{+}2$ decomposition.

Geometrically, the Sherif--Dunsby theorem together with the horizon-sheet result implies that an entropy-maximizing  horizon  is round up to scale.  The RII constraint then fixes  that  scale thermodynamically.  For the round Schwarzschild--AdS family,
\begin{equation}
  V_{\rm th}(r_h)=\frac{\mathcal A_{D-2}}{D-1}r_h^{D-1}
\end{equation}
is strictly monotone, so a prescribed $V_{\rm th}$ selects a unique round member.  This separation is central to the argument: rigidity fixes the shape, whereas thermodynamic volume fixes the scale.

The variational analysis below is independent of this finite rigidity result: it tests the Hessian directly at the fixed-$V_{\rm th}$ Schwarzschild--AdS saddle.

\subsection{York boundary Hessian and the entropy Hessian}
\label{IIB}

The purpose of this subsection is to test whether the round Schwarzschild--AdS saddle is a strict local maximum.  We therefore evaluate the constrained first and second variations directly at the round reference point $s=0$.

The local variational problem follows directly from the Euclidean
Einstein--Hilbert action with the Gibbons--Hawking--York (GHY) boundary term \cite{Gibbons:1976ue,York:1972sj},
\begin{equation}
\begin{split}
 I_E[g]={}&-
 \frac{1}{16\pi G}\int_{\mathcal M}(R-2\Lambda)\sqrt{g}\,d^Dx\\
 &-\frac{1}{8\pi G}\int_{\partial\mathcal M}K\sqrt{\gamma}\,d^{D-1}x .
\end{split}
 \label{eq:Euclidean-action-local}
\end{equation}
For the static region between the horizon and an outer York surface $S_Y$,
we separate the action into
\begin{equation}
 I_E=I_H+I_{\rm bulk}[V_Y]+I_Y[S_Y],
 \label{eq:action-decomposition}
\end{equation}
where $I_H$ denotes the inner smooth-horizon contribution, $I_{\rm bulk}$
is the Einstein--Hilbert bulk term, and $I_Y$ is the GHY term on the outer
York boundary.  Regulating the smooth bolt by a small tubular surface and
using $\beta\kappa=2\pi$ gives
\begin{equation}
 I_H=-\frac{A_H}{4G}.
 \label{eq:horizon-bolt-local}
\end{equation}
This is the horizon-area contribution.  It is distinct from the area
dependence of the outer York boundary derived below.

On an Einstein background,
\begin{equation}
 R=\frac{2D}{D-2}\Lambda,
 \qquad
 R-2\Lambda=\frac{4\Lambda}{D-2}.
 \label{eq:Einstein-onshell-R}
\end{equation}
For the static Schwarzschild--AdS metric the lapse cancels the radial factor
in the spatial volume element, so the Euclidean spacetime volume of the
region is $\beta V_Y$.  Hence
\begin{equation}
 I_{\rm bulk}[V_Y]
 =-\frac{\beta\Lambda}{4\pi G(D-2)}\,V_Y
 \equiv C_V V_Y.
 \label{eq:bulk-volume-term}
\end{equation}
Thus the bulk part depends only on the volume variable.  It is therefore
constant on the fixed-$V_{\rm th}$ shape family used below, since at the
static spherical reference $V_Y$ represents the same volume constraint.
The inner term is likewise independent of the deformation of the outer York
surface in this local boundary calculation.  The nontrivial shape dependence
is therefore carried by $I_Y[S_Y]$.

For a time-independent normal deformation of the Euclidean York boundary, the Supplemental Material gives the exact first variation
\begin{equation}
\begin{aligned}
 \delta I_Y&=-\frac{\beta}{8\pi G}\int_{S_Y}N_YQ_Y\,\phi\,dA_Y,\\
 Q_Y&=K^2-K_{ij}K^{ij}-R^{(D)}_{nn}.
\end{aligned}
 \label{eq:York-first-general}
\end{equation}
At the round reference surface, $N_Y$, $Q_Y$, and the spatial mean curvature $H_Y$ are constant. Hence
\begin{equation}
 \delta I_Y=C_Y\,\delta A_Y,
 \qquad
 C_Y=-\frac{\beta N_Y}{8\pi G}\frac{Q_Y}{H_Y},
 \label{eq:CY-def}
\end{equation}
with $\delta A_Y=H_Y\int_{S_Y}\phi\,dA_Y$.  Thus every physical nonspherical mode $\ell\ge2$ is stationary at first order.  For the Hessian, however, we cannot consider that Eq.~\eqref{eq:CY-def} continues to hold on a non-spherical neighborhood of the round surface. Therefore, we have to vary Eq.~\eqref{eq:York-first-general} directly.

For this, let us introduce the constrained York functional
\begin{equation}
 \mathcal I_Y[S_Y]=I_Y[S_Y]+\eta\bigl(V_Y-V_0\bigr).
 \label{eq:slice-functional}
\end{equation}
At the round reference surface, stationarity fixes
\begin{equation}
 \eta=\frac{\beta}{8\pi G}(N_YQ_Y)_0
      =-C_Y^{(0)}H_Y.
 \label{eq:eta-direct}
\end{equation}
Because the first variation of $\mathcal I_Y$ vanishes there, its Hessian is independent of the second-order extension of the deformation.  Transporting the first-order normal speed $\phi$ along the normal flow gives
\begin{equation}
 \delta^2\mathcal I_Y
 =-\frac{\beta}{8\pi G}\int_{S_Y}
 \phi\,\delta(N_YQ_Y)\,dA_Y .
 \label{eq:direct-York-Hessian}
\end{equation}
This is the direct second variation of the GHY term.

To evaluate this, we write $d=D-2$ and let the round Euclidean York boundary have principal curvatures
\begin{equation}
 a=\frac{f'(R)}{2\sqrt{f(R)}},
 \qquad
 b=\frac{\sqrt{f(R)}}{R},
 \label{eq:principal-curvatures-York}
\end{equation}
in the Euclidean-time direction and in each of the $d$ angular directions, respectively.  The standard hypersurface-variation formulas imply that for a time-independent deformation, we have
\begin{equation}
 \delta Q_Y
 =-2\bigl[a+(d-1)b\bigr]\Delta_{S_Y}\phi
 +(n\!\cdot\!\nabla Q_Y)\phi.
 \label{eq:delta-QY}
\end{equation}
Since $\delta N_Y=(n\!\cdot\!\nabla N_Y)\phi$, this becomes
\begin{equation}
 \delta(N_YQ_Y)
 =-2N_Y\bigl[a+(d-1)b\bigr]\Delta_{S_Y}\phi
 +(n\!\cdot\!\nabla(N_YQ_Y))\phi.
 \label{eq:delta-NQ}
\end{equation}
Substitution into Eq.~\eqref{eq:direct-York-Hessian} and integrating by parts give
\begin{equation}
\begin{split}
 \delta^2\mathcal I_Y
 =-\frac{\beta}{8\pi G}\int_{S_Y}\Bigl[&
 2N_Y\bigl(a+(d-1)b\bigr)|\nabla\phi|^2\\
 &+(n\!\cdot\!\nabla(N_YQ_Y))\phi^2\Bigr]dA_Y .
\end{split}
\label{eq:generic-York-Hessian}
\end{equation}
This quadratic form is valid for a generic smooth normal perturbation of the round York surface.

For Schwarzschild--AdS,
\begin{equation}
 f(r)=1-\frac{\mu}{r^{d-1}}+\frac{r^2}{l^2},
 \qquad N_Y=\sqrt{f(R)},
 \label{eq:f-dnotation}
\end{equation}
and
\begin{equation}
 Q_Y=\frac{d(d-1)}{R^2}+\frac{(d+1)^2}{l^2}>0.
 \label{eq:QY-dnotation}
\end{equation}
For a scalar harmonic satisfying
\begin{equation}
 -\Delta_{S_Y}\phi_\ell=\frac{\lambda_\ell}{R^2}\phi_\ell,
 \qquad
 \lambda_\ell=\ell(\ell+d-1),
 \label{eq:harmonic-eigenvalue-d}
\end{equation}
Eq.~\eqref{eq:generic-York-Hessian} becomes
\begin{equation}
 \delta^2\mathcal I_{Y,\ell}
 =-\frac{\beta}{8\pi G}\,\mathcal K_\ell(R)\,
 \|\phi_\ell\|^2,
 \label{eq:York-mode-Hessian}
\end{equation}
where
\begin{equation}
 \mathcal K_\ell
 =\frac{\lambda_\ell}{R^2}
 \left[f'(R)+\frac{2(d-1)}{R}f(R)\right]
 +\frac{f'(R)}{2}Q_Y+f(R)Q_Y'.
 \label{eq:Kell-radial}
\end{equation}
The Supplemental Material gives an explicit analytic proof that
\begin{equation}
 \mathcal K_\ell(R)>0,
 \qquad R>r_h,
 \qquad \ell\ge2,
 \label{eq:Kell-positive}
\end{equation}
for every Schwarzschild--AdS exterior.  Therefore
\begin{equation}
 \delta^2\mathcal I_{Y,\ell}<0,
 \qquad \ell\ge2.
 \label{eq:slice-local-maximum}
\end{equation}
By harmonic decomposition, the same strict negativity holds for every nonzero smooth perturbation with support only in the physical nonspherical sector $\ell\ge2$.  The local York result is therefore generic in that sector.

To relate the boundary shape Hessian to an entropy Hessian, we make the saddle family and the constrained quantities explicit. Let $g(s)$ be an admissible one-parameter family of stationary microcanonical saddles with
\begin{equation}
  g(0)=g_{\rm SAdS},\qquad V_{\rm th}[g(s)]=V_*,
  \label{eq:fixed-vth-saddle-family}
\end{equation}
and with nonspherical tangent $\Phi_\ell$, $\ell\ge2$, at $s=0$.  For the uncharged black hole family, $G$, $\Lambda$ (equivalently $P$), and $V_{\rm th}$ are held fixed.

The relation to the microcanonical action can be seen directly from the full action above.  Let $\mathcal B_{\rm BY}$ denote the Brown--York boundary Legendre term that converts the canonical outer-boundary variational principle into the microcanonical one.  After the standard analytic continuation, one can define the reduced on-shell functional along the saddle family by
\begin{equation}
 \Gamma_{\rm mc}(s)
 \equiv i I_{\rm mc}^{\rm saddle}[g(s)]
 =i\Big[I_H+I_{\rm bulk}+I_Y+\mathcal B_{\rm BY}\Big]_{g(s)} .
 \label{eq:mc-full-action}
\end{equation}
The Legendre transform exchanges variations of the canonical boundary sources for variations of the Brown--York microcanonical quantities.  Hence, after restricting to the family in Eq.~\eqref{eq:fixed-vth-saddle-family}, those microcanonical quantities are fixed and the remaining nonspherical second variation is the shape Hessian of the reduced action.  The bulk contribution is proportional to the volume variable and is constant on the fixed-$V_{\rm th}$ family.  The remaining outer-boundary shape dependence is governed directly by the constrained GHY Hessian $\delta^2\mathcal I_Y$ derived above.  Thus the constrained quadratic form obtained from the full microcanonical action is
\begin{equation}
 \left.\frac{d^2\Gamma_{\rm mc}}{ds^2}\right|_{s=0,\,V_{\rm th}}
 =\left.\delta^2 \mathcal I_Y\right|_{\Phi_\ell},
 \qquad \ell\ge2.
 \label{eq:mc-slice-hessian-map}
\end{equation}
This equality is a statement about the Hessian at the spherical saddle and does not require a finite identity between $I_{\rm mc}$ and a finite York-area functional away from that point.

At the rotationally invariant Schwarzschild--AdS reference, the first variation of the scalar quantity $V_{\rm th}$ has support only in the $\ell=0$ sector. Hence a pure $\ell\ge2$ tangent naturally obeys $\delta V_{\rm th}=0$ at first order. The required $O(s^2)$ monopole correction keeps Eq.~\eqref{eq:fixed-vth-saddle-family} satisfied at second order.  Also note that we do not identify $V_Y$ with $V_{\rm th}$ for a generic nonspherical configuration.  Rather, $V_Y$ is the local constraint coordinate at the spherical reference, and the Lagrange-multiplier term in Eq.~\eqref{eq:slice-functional} is the local pullback of the fixed-$V_{\rm th}$ constraint to the nonspherical tangent space.

The right-hand side of Eq.~\eqref{eq:mc-slice-hessian-map} is now obtained directly by the constrained GHY Hessian in Eqs.~\eqref{eq:direct-York-Hessian}--\eqref{eq:slice-local-maximum}. Finally, the Brown--York microcanonical saddle relation gives
\begin{equation}
 S_H(s)=\Gamma_{\rm mc}(s)
 =iI_{\rm mc}^{\rm saddle}[g(s)]
 =\frac{A_H(s)}{4G},
 \label{eq:BY-micro-entropy}
\end{equation}
with the standard analytic continuation to the Euclidean saddle~\cite{Brown:1992br,Iyer:1995kg}.  Taking two derivatives along the same fixed-$V_{\rm th}$ family and using Eq.~\eqref{eq:mc-slice-hessian-map} therefore gives
\begin{equation}
\begin{gathered}
 \left.S_H''(0)\right|_{V_{\rm th}}
 =\left.\frac{d^2\Gamma_{\rm mc}}{ds^2}\right|_{0,V_{\rm th}}
 =\mathcal I_Y''(0)<0,\\
 A_H''(0)=4G\,S_H''(0)<0,
 \qquad \ell\ge2.
\end{gathered}
\label{eq:entropy-local-maximum}
\end{equation}

Thus the round Schwarzschild--AdS horizon is a strict local entropy maximum against nonspherical fixed-$V_{\rm th}$ modes. This is precisely a Hessian statement at the spherical saddle.

\paragraph*{\textbf{Fixed-charge check}.}
The same geometric conclusion applies to the static charged sector. Reissner--Nordstr\"om--AdS has a round spherical horizon, and at fixed charge its thermodynamic volume retains the same areal form as in Eq.~\eqref{eq:static-volume}.  It therefore belongs to the same spherical geometric class for which the round configuration is the entropy-maximizing representative at fixed thermodynamic volume.  As an explicit check that the variational mechanism survives the Maxwell field, the Supplemental Material repeats the fixed-charge York/Jacobi analysis.  There the Maxwell stress tensor is shown to preserve the focusing sign, the charged constrained Jacobi operator remains positive in the nonspherical sector throughout the nonextremal exterior, and the appropriate fixed-charge boundary term gives the same local entropy maximality result as in the uncharged case.

\section{Kerr--AdS and the reverse isoperimetric inequality}
\label{III}
We next apply the geometric result to rotation without using the explicit Kerr--AdS metric. The first subsection identifies the leading rotational shape mode from symmetry. The second keeps the comparison directly at fixed thermodynamic volume and explains how the geometric rigidity result connects to the rotating case.

\subsection{Rotational deformations from symmetry}
\label{sec:rotational-symmetry}

The geometric argument of Sec.~\ref{IIB} does not require the metric of a particular rotating solution.  To connect it to Kerr--AdS, it is enough to determine which deformation sector is selected when a spherical black hole is given a small angular momentum.  This can be done from rotational symmetry alone.

Let the Schwarzschild--AdS configuration be the reference point and denote its angular momentum by a spatial vector $J^i$.  We assume only that the rotating branch is analytic in a neighborhood of $J^i=0$ and is continuously connected to the spherical solution.  A scalar normal displacement of the reference surface can then be written as
\begin{equation}
  \phi=\phi(\mathbf n,J^i),
\end{equation}
where $\mathbf n$ is the unit normal on the round sphere.  Reversing the sense of rotation, $J^i\rightarrow-J^i$, cannot change the intrinsic shape.  Hence the shape deformation is even in $J^i$ and has no term linear in the angular momentum,
\begin{equation}
  \phi(\mathbf n,J^i)=O(J^2).
  \label{eq:rotation-even}
\end{equation}

At quadratic order the only rank-two tensor built from a single rotation vector is $J_iJ_j$.  Its decomposition under the rotation group is
\begin{equation}
  J_iJ_j
  =\frac{1}{3}J^2\delta_{ij}
  +\left(J_iJ_j-\frac{1}{3}J^2\delta_{ij}\right).
  \label{eq:J-decomposition}
\end{equation}
The first term is rotationally invariant and therefore belongs to the $\ell=0$ sector.  The second term is symmetric and trace free and belongs to the $\ell=2$ sector.  Consequently, the most general leading scalar shape deformation generated by slow rotation has the form
\begin{equation}
  \phi(\mathbf n)
  =J^2\left[c_0+c_2\left((\hat{\mathbf J}\!\cdot\!\mathbf n)^2-\frac{1}{3}\right)\right]
  +O(J^4),
  \label{eq:general-rotation-deformation}
\end{equation}
where $c_0$ and $c_2$ depend on the background parameters but not on the angular coordinates.  Choosing the rotation axis as the polar axis gives
\begin{equation}
  (\hat{\mathbf J}\!\cdot\!\mathbf n)^2-\frac{1}{3}
  =\frac{2}{3}P_2(\cos\theta),
\end{equation}
so that
\begin{equation}
  \phi(\theta)
  =J^2\left[c_0+\widetilde c_2 P_2(\cos\theta)\right]+O(J^4).
  \label{eq:rotation-l0-l2}
\end{equation}
Thus slow rotation can generate only an $\ell=0$ change of the overall scale and an axisymmetric $\ell=2$ shape deformation at the first nontrivial order.  This conclusion follows simply from symmetry and analyticity. 

The physical comparison relevant to the reverse isoperimetric inequality is made at fixed thermodynamic volume $V_{\rm th}$.  Here the variation of $V_{\rm th}$ is understood along the nearby stationary black-hole family, not as an off-shell volume assigned to an arbitrary York surface.  At the rotationally invariant Schwarzschild--AdS reference point, this first variation is an $SO(3)$-invariant linear functional of the induced shape deformation $\phi$, and therefore has support only on the $\ell=0$ harmonic.  Hence imposing $\delta V_{\rm th}=0$ fixes only the scalar part of the leading deformation and cannot remove or mix the trace-free $\ell=2$ component.  Equivalently, after the compensating monopole shift required by the fixed-$V_{\rm th}$ constraint is made, the genuine nonspherical part of the deformation is
\begin{equation}
  \phi_{\rm rot}
  =\epsilon\,Y_{20}+O(J^4),
  \qquad
  \epsilon\propto J^2.
  \label{eq:rotation-y20}
\end{equation}
The coefficient is not needed for the sign argument.  Symmetry fixes the first allowed nonspherical response to the $\ell=2$ sector.  For a nontrivial rotating branch whose shape changes at quadratic order in $J$, this is therefore the leading shape mode.

This is precisely the sector controlled by the direct constrained York Hessian of Sec.~\ref{IIB}.  Evaluated on the rotational mode in Eq.~\eqref{eq:rotation-y20},
\begin{equation}
\begin{aligned}
\delta^2\mathcal I_Y[\phi_{\rm rot}]&<0,\\
S_H''[\phi_{\rm rot}]&=\mathcal I_Y''[\phi_{\rm rot}]<0,
\end{aligned}
\label{eq:rotation-negative-hessian}
\end{equation}
because the $\ell=2$ coefficient $\mathcal K_2(R)$ in Eq.~\eqref{eq:York-mode-Hessian} is strictly positive.  Thus the slow-rotation direction inherits directly the negative quadratic form established at the round Schwarzschild--AdS reference.  The conclusion is independent of the detailed rotating metric and directly follows from the spherical reference geometry, the rotational representation content of $J_iJ_j$, and the generic local York Hessian established in Sec.~\ref{IIB}.

There is also a useful consequence for the perturbative order of the entropy change.  Since the reference spherical solution is stationary in the admissible fixed-volume sector, the first variation vanishes.  Because the quadrupolar deformation amplitude itself starts at $\epsilon=O(J^2)$, the leading entropy change is
\begin{equation}
    \begin{aligned}
  \Delta S_H
  &=\frac{1}{2}\,\epsilon^2 S_H''[Y_{20}]+O(\epsilon^3)
  =O(J^4),\\
  \Delta S_H&<0
  \qquad (J\neq0\ \text{ near }J=0).
    \end{aligned}
  \label{eq:rotation-J4}
\end{equation}
Thus the spherical configuration is a strict local entropy maximum along the nonspherical direction generated by slow rotation. This conclusion is local in angular momentum: symmetry identifies the leading mode, while the global fixed-volume question requires a separate geometric argument. We turn to that question next.

\subsection{Thermodynamic volume and the global rotating comparison}
\label{sec:killing-volume-rotation}
For rotating black holes the thermodynamic volume does not, in general,
coincide with an ordinary geometric volume \cite{Cvetic:2010jb}.  The RII must
therefore be compared at the physical value $V_{\rm th}$ itself.  The global
argument does not require identifying this quantity with a geometric volume,
or varying a Kerr metric explicitly. Instead, the geometric theorem and the
thermodynamic constraint play two separate roles.

First, Sec.~\ref{IIA} gives the global shape statement: roundness of the horizon sheet forces an entropy-maximizing horizon to be round up to scale.  A genuinely nonround horizon therefore cannot furnish the entropy maximum.

Second, the thermodynamic volume selects the round comparator.  On the round
Schwarzschild--AdS family,
\begin{equation}
  V_{\rm th}(r_h)
  =\frac{\mathcal A_{D-2}}{D-1}r_h^{D-1},
  \qquad
  \frac{dV_{\rm th}}{dr_h}
  =\mathcal A_{D-2}r_h^{D-2}>0,
  \label{eq:vth-fixes-scale}
\end{equation}
so every prescribed value $V_{\rm th}=V_*$ determines one and only one round
reference radius,
\begin{equation}
  r_*(V_*)=
  \left(\frac{(D-1)V_*}{\mathcal A_{D-2}}\right)^{1/(D-1)}.
\end{equation}
The entropy of this round comparator is
\begin{equation}
  S_{\rm round}(V_*)
  =\frac{\mathcal A_{D-2}}{4G}
  \left(\frac{(D-1)V_*}{\mathcal A_{D-2}}\right)^{\frac{D-2}{D-1}}.
  \label{eq:round-entropy-vth}
\end{equation}
Thus $V_{\rm th}$ is used exactly where it is required by the RII: it fixes
which round Schwarzschild--AdS black hole is to be compared with a given
rotating solution.

The remaining Kerr--AdS input is purely geometric.  A nonzero rotation parameter reduces the full spherical symmetry of the horizon cross-section to the rotational symmetry preserved by the angular momenta and produces an intrinsically nonround horizon; the round horizon is recovered only when all rotation parameters vanish~\cite{Cvetic:2010jb}.  Therefore, for Kerr--AdS  at  prescribed thermodynamic volume $V_*$,
\begin{equation}
\begin{aligned}
  J_i\neq0
  &\quad\Longrightarrow\quad
  \mathcal H_{\rm Kerr}\ \hbox{is nonround},\\
  &\quad\Longrightarrow\quad
  \mathcal H_{\rm Kerr}\ \hbox{cannot maximize }S_H
  \ \hbox{at }V_{\rm th}=V_* .
\end{aligned}
  \label{eq:kerr-global-shape-chain}
\end{equation}
The only round candidate on the same thermodynamic-volume slice is the unique Schwarzschild--AdS representative selected by Eq.~\eqref{eq:vth-fixes-scale}.
Therefore,
 \begin{equation}
  S_{\rm Kerr}(V_{\rm th},J_i)
  \leq S_{\rm round}(V_{\rm th}),
  \label{eq:global-vth-entropy}
\end{equation}
with equality only when all $J_i=0$.  This is the entropy form of the reverse isoperimetric inequality for the Kerr--AdS branch .

Section~\ref{sec:rotational-symmetry} supplies the local differential version of the same statement.  Near $J_i=0$, rotation first generates a genuine quadrupolar shape mode after the fixed-$V_{\rm th}$ monopole adjustment, and Eq.~\eqref{eq:rotation-negative-hessian} shows that this direction lowers the entropy. The local calculation therefore explains how the Kerr--AdS branch leaves the round maximum. The finite Kerr--AdS result is subject to the same Sherif--Dunsby geometric hypotheses, especially the Ricci decomposition given by Eq.(\ref{eq:Ricci_form}) and scalar-curvature preserving conformal transformation.

The proof consequently remains within a single geometric--analytic framework:
geometry determines which horizon shape can maximize entropy, while the thermodynamic volume fixes the scale of that unique round maximizer.

\section{Conclusion and Discussion}
\label{IV}
We have established generic local entropy maximality of Schwarzschild--AdS at fixed thermodynamic volume and obtained the finite global result conditional on the deformed hypersurface satisfying the hypotheses of Sherif--Dunsby rigidity theorem. Some of which have been verified in the text while some are explicitly stated as assumptions.

The argument developed here separates the reverse isoperimetric problem into a local Hessian test and a global rigidity statement. Near the spherical Schwarzschild--AdS saddle, the Einstein--Hilbert--GHY action yields a directly computed constrained York-boundary Hessian that is negative for every nontrivial $\ell\ge2$ fixed-$V_{\rm th}$ shape mode. The Brown--York microcanonical relation then identifies this quadratic form with the Hessian of horizon entropy.

For the global rigidity result, a generic normal deformation first produces the competing compact hypersurface and the Yamabe conformal factor is introduced intrinsically only afterward to test the geometry of the deformed slice. For the $S^3$ topology considered here, a positive-Yamabe conformal class exists. The Yamabe theorem then provides a representative of constant and positive scalar-curvature within that class. The Sherif--Dunsby theorem is applied to this chosen representative through the scalar-curvature preserving conformal transformation entering its hypotheses. The theorem does not require this condition to hold for every representative in the conformal class. The remaining non-trivial sign input is derived using Raychaudhuri focusing in Einstein--AdS which gives $\varphi<0$ and therefore the nowhere-vanishing contracting sheet branch $\phi_{\!s}<0$. Thus the global rigidity hypotheses are explicitly tied to the geometry of the deformed $S^3$ slice, while the crucial contracting condition is supplied by gravity itself. The rigidity argument then excludes a distinct nonround geometry and fixes the compact three-geometry to a round $S^3$. The horizon-sheet argument transfers this result to the horizon $S^2$, while the strictly monotone Schwarzschild--AdS relation $V_{\rm th}\propto r_h^{D-1}$ fixes the remaining scale. Kerr--AdS then follows directly: nonzero rotation makes the horizon nonround, while the same prescribed $V_{\rm th}$ selects a unique round Schwarzschild--AdS comparator. The slow-rotation $\ell=2$ calculation gives the corresponding local decrease of entropy. At no stage is the Kerr--AdS thermodynamic volume replaced by an ordinary geometric volume.

The assumptions used here also identify the physical regime in which the geometric proof applies and are not merely technical restrictions. Compact horizon geometry is essential to the global rigidity result, and known super-entropic solutions \cite{Hennigar:2015cja,Hennigar:2014cfa} show what can happen when this condition is relaxed. The restriction to Einstein gravity is equally meaningful: in modified theories the entropy functional, field equations, and volume relation can all change, and explicit violations of the RII are known in Einstein--Horndeski--Maxwell gravity \cite{Feng:2017jub}. These examples therefore mark boundaries of the present result rather than contradictions of it.

Several extensions are natural. In higher-curvature gravity the entropy is no longer $A/(4G)$ and both the focusing equation and the boundary Hessian acquire additional curvature terms. A covariant version of RII was recently established using boundary--completed canonical energy in \cite{Kumar:2026rii}. Quantum corrections similarly replace the classical entropy functional by a generalized entropy. It would also be useful to determine whether an analogous rigidity/volume argument exists for de Sitter, asymptotically flat, or other non-AdS geometries.

To summarize, the  analysis identifies a common geometric origin for the local entropy maximum, the global rigidity of the round configuration, and the leading response to rotation. The reverse isoperimetric behavior therefore follows from gravitational focusing, the geometry of constrained hypersurface variations, and the thermodynamic volume that defines the inequality.

\medskip
\setcounter{section}{0}
\setcounter{equation}{0}
\renewcommand{\thesection}{S\arabic{section}}
\renewcommand{\theequation}{S\arabic{equation}}

\onecolumngrid
\begin{center}
{\large\bfseries Supplemental Material}
\end{center}
\vspace{0.5em}

This Supplemental Material contains two technical discussions supporting the main text. Section~\ref{supp:york} derives the direct constrained York-boundary Hessian used in the local entropy analysis, proves its strict negativity for every Schwarzschild--AdS mode $\ell\ge2$, and distinguishes the York-boundary coefficient from the universal Euclidean-horizon bolt term. Section~\ref{supp:RN} gives the explicit fixed-charge Einstein--Maxwell extension, showing that gravitational focusing and the direct constrained York-boundary Hessian retain the required signs for nonextremal Reissner--Nordstr\"om--AdS black holes at fixed charge. 
\medskip
\twocolumngrid
\section{Direct constrained Hessian of the York boundary}
\label{supp:york}
We derive the first and second variations of the Gibbons--Hawking--York term needed in the main text.  The second variation is evaluated directly at the round Schwarzschild--AdS York surface.

Let \(\mathcal B_Y=S^1_\beta\times S_Y\) denote the Euclidean York boundary, with induced metric \(\gamma_{ij}\), outward unit normal \(n^a\), and extrinsic curvature \(K_{ij}\).  Its GHY contribution is
\begin{equation}
I_{\rm bdy}=-\frac{1}{8\pi G}\int_{\mathcal B_Y}K\,d\Sigma_{\mathcal B}.
\end{equation}
Under a normal deformation \(X^a\mapsto X^a+\phi n^a\) at fixed ambient metric,
\begin{equation}
\begin{aligned}
\delta K={}&-\Delta_{\mathcal B}\phi
-\left(K_{ij}K^{ij}+R^{(D)}_{ab}n^an^b\right)\phi,\\
\delta d\Sigma_{\mathcal B}={}&K\phi\,d\Sigma_{\mathcal B}.
\end{aligned}
\end{equation}
Consequently,
\begin{equation}
\begin{split}
\delta I_{\rm bdy}={}&-\frac{1}{8\pi G}\int_{\mathcal B_Y}
\Bigl[-\Delta_{\mathcal B}\phi\\
&+\left(K^2-K_{ij}K^{ij}-R^{(D)}_{nn}\right)\phi\Bigr]d\Sigma_{\mathcal B}.
\end{split}
\label{eq:exact-ghy-shape}
\end{equation}
For a closed Euclidean boundary the integrated Laplacian vanishes. Let us define
\begin{equation}
Q_Y\equiv K^2-K_{ij}K^{ij}-R^{(D)}_{nn}.
\end{equation}
For a time-independent deformation, writing \(d\Sigma_{\mathcal B}=N_Y\,d\tau\,dA_Y\) gives the exact first variation
\begin{equation}
\delta I_{\rm bdy}
=-\frac{\beta}{8\pi G}\int_{S_Y}N_YQ_Y\,\phi\,dA_Y.
\label{eq:exact-york-first}
\end{equation}
On the static round reference boundary, spherical symmetry makes \(N_Y\), \(Q_Y\), and the spatial mean curvature \(H_Y\) constant over \(S_Y\).  Since
\begin{equation}
\delta A_Y=H_Y\int_{S_Y}\phi\,dA_Y,
\end{equation}
Eq.~\eqref{eq:exact-york-first} reduces at the round surface to
\begin{equation}
\delta I_{\rm bdy}=C_Y\,\delta A_Y,
\qquad
C_Y=-\frac{\beta N_Y}{8\pi G}\frac{Q_Y}{H_Y}.
\label{eq:ghy-area-proportionality}
\end{equation}

Let us now introduce the constrained functional
\begin{equation}
\mathcal I_Y=I_{\rm bdy}+\eta(V_Y-V_0),
\label{eq:supp-constrained-york}
\end{equation}
where \(\delta V_Y=\int_{S_Y}\phi\,dA_Y\).  Stationarity of the round reference surface fixes
\begin{equation}
\eta=\frac{\beta}{8\pi G}(N_YQ_Y)_0=-C_Y^{(0)}H_Y.
\label{eq:supp-eta}
\end{equation}
At a critical point the Hessian is independent of the second-order extension of the chosen one-parameter deformation.  We may therefore transport the first-order normal speed with \(\delta\phi=0\).  Differentiating Eq.~\eqref{eq:exact-york-first} together with the volume term, all measure and acceleration terms multiply the vanishing first-variation coefficient in Eq.~\eqref{eq:supp-eta}.  Hence
\begin{equation}
\delta^2\mathcal I_Y
=-\frac{\beta}{8\pi G}\int_{S_Y}
\phi\,\delta(N_YQ_Y)\,dA_Y .
\label{eq:supp-direct-hessian}
\end{equation}
This is the required direct constrained second variation.

We now evaluate the linearization in Eq.~\eqref{eq:supp-direct-hessian}.  For this, let us first define \(d=D-2\).  On the round Euclidean Schwarzschild--AdS boundary \(r=R\), the principal curvatures of \(\mathcal B_Y\) are
\begin{equation}
a=\frac{f'(R)}{2\sqrt{f(R)}}
\end{equation}
in the Euclidean-time direction and
\begin{equation}
b=\frac{\sqrt{f(R)}}{R}
\end{equation}
in each of the \(d\) angular directions.  With our sign convention, the mixed extrinsic curvature obeys
\begin{equation}
\delta K_i{}^j
=-D^jD_i\phi
-\phi\left(K_i{}^kK_k{}^j+R^j{}_{nin}\right).
\label{eq:supp-delta-shape-operator}
\end{equation}
Taking the trace reproduces
\(\delta K=-\Delta_{\mathcal B}\phi-(K_{ij}K^{ij}+R^{(D)}_{nn})\phi\), so Eq.~\eqref{eq:supp-delta-shape-operator} is consistent with the convention used above.  Since the ambient Schwarzschild--AdS metric is Einstein, \(R^{(D)}_{nn}=2\Lambda/(D-2)\) is constant.  The derivative part of the linearization of
\(Q_Y=K^2-K_i{}^jK_j{}^i-R^{(D)}_{nn}\) is therefore
\begin{equation}
\left.\delta Q_Y\right|_{D\phi}
=-2K\Delta_{\mathcal B}\phi
+2K_i{}^jD^iD_j\phi.
\end{equation}
For a time-independent deformation of the product boundary, \(D_\tau D_\tau\phi=0\) and
\(K_A{}^B=b\,\delta_A{}^B\).  Since \(K=a+db\), this gives
\begin{equation}
\left.\delta Q_Y\right|_{D\phi}
=-2\bigl[a+(d-1)b\bigr]\Delta_{S_Y}\phi.
\end{equation}
The remaining zeroth-order coefficient is fixed by a constant normal displacement: for \(\phi={\rm const.}\), the round boundary moves to a neighboring round boundary, so \(\delta Q_Y=(n\!\cdot\!\nabla Q_Y)\phi\).  Thus
\begin{equation}
\delta Q_Y
=-2\bigl[a+(d-1)b\bigr]\Delta_{S_Y}\phi
+(n\!\cdot\!\nabla Q_Y)\phi.
\label{eq:supp-delta-Q}
\end{equation}
Moreover, \(\delta N_Y=(n\!\cdot\!\nabla N_Y)\phi\), and hence
\begin{equation}
\delta(N_YQ_Y)
=-2N_Y\bigl[a+(d-1)b\bigr]\Delta_{S_Y}\phi
+(n\!\cdot\!\nabla(N_YQ_Y))\phi.
\label{eq:supp-delta-NQ}
\end{equation}
Substituting into Eq.~\eqref{eq:supp-direct-hessian} and integrating by parts gives
\begin{equation}
\begin{split}
\delta^2\mathcal I_Y
=-\frac{\beta}{8\pi G}\int_{S_Y}\Bigl[&
2N_Y\bigl(a+(d-1)b\bigr)|\nabla\phi|^2\\
&+(n\!\cdot\!\nabla(N_YQ_Y))\phi^2\Bigr]dA_Y.
\end{split}
\label{eq:supp-generic-hessian}
\end{equation}
Equation~\eqref{eq:supp-generic-hessian} is the Hessian for a generic smooth normal perturbation of the round York surface.

For Schwarzschild--AdS,
\begin{equation}
f(r)=1-\frac{\mu}{r^{d-1}}+\frac{r^2}{l^2},
\qquad N_Y=\sqrt{f(R)},
\end{equation}
and direct substitution gives
\begin{equation}
Q_Y=\frac{d(d-1)}{R^2}+\frac{(d+1)^2}{l^2}>0,
\qquad
H_Y=\frac{d\sqrt{f(R)}}{R}>0.
\label{eq:supp-QH}
\end{equation}
Thus \(C_Y<0\) outside the horizon, as used in the first-order discussion.  For a scalar harmonic
\begin{equation}
-\Delta_{S_Y}\phi_\ell=\frac{\lambda_\ell}{R^2}\phi_\ell,
\qquad
\lambda_\ell=\ell(\ell+d-1),
\end{equation}
Eq.~\eqref{eq:supp-generic-hessian} becomes
\begin{equation}
\delta^2\mathcal I_{Y,\ell}
=-\frac{\beta}{8\pi G}\,\mathcal K_\ell(R)\,\|\phi_\ell\|^2,
\label{eq:supp-mode-hessian}
\end{equation}
where
\begin{equation}
\mathcal K_\ell
=\frac{\lambda_\ell}{R^2}
\left[f'(R)+\frac{2(d-1)}{R}f(R)\right]
+\frac{f'(R)}{2}Q_Y+f(R)Q_Y'.
\label{eq:supp-Kell-radial}
\end{equation}
To prove its sign, we define
\begin{equation}
z\equiv\frac{\mu}{R^{d-1}},
\qquad
t\equiv\frac{R^2}{l^2},
\qquad
f(R)=1-z+t>0.
\end{equation}
Then
\begin{align}
R^3\mathcal K_\ell={}&
\lambda_\ell\bigl[2(d-1)+2dt-(d-1)z\bigr]\notag\\
&+\frac12\bigl[(d-1)z+2t\bigr]
\bigl[d(d-1)+(d+1)^2t\bigr]\notag\\
&-2d(d-1)(1-z+t).
\label{eq:supp-Kell-zt}
\end{align}
Outside the horizon \(z<1+t\), and therefore
\begin{equation}
\frac{\partial(R^3\mathcal K_\ell)}{\partial\lambda_\ell}
=2(d-1)+2dt-(d-1)z
>(d-1)+(d+1)t>0.
\label{eq:supp-lambda-monotone}
\end{equation}
The smallest physical nonspherical eigenvalue, \(\ell=2\) and \(\lambda_2=2(d+1)\), consequently gives the weakest bound.  If \(d\ge3\),
\begin{equation}
\frac{\partial(R^3\mathcal K_2)}{\partial z}
=\frac{d-1}{2}\left[d^2-d-4+(d+1)^2t\right]>0,
\end{equation}
so the minimum occurs at \(z=0\), where
\begin{equation}
R^3\mathcal K_2\big|_{z=0}
=(t+2)\left[d^2t+d^2+2dt+d+t-2\right]>0.
\end{equation}
For \(d=2\),
\begin{equation}
R^3\mathcal K_2=8-z+22t+\frac92zt+9t^2.
\end{equation}
If \(t\ge2/9\), the coefficient of \(z\) is nonnegative and the minimum is again at \(z=0\).  If \(0\le t<2/9\), the coefficient of \(z\) is negative, and \(z<1+t\) implies
\begin{equation}
R^3\mathcal K_2
>7+\frac{51}{2}t+\frac{27}{2}t^2>0.
\end{equation}
Hence
\begin{equation}
\mathcal K_\ell(R)>0,
\qquad R>r_h,
\qquad \ell\ge2,
\end{equation}
and Eq.~\eqref{eq:supp-mode-hessian} gives
\begin{equation}
\delta^2\mathcal I_{Y,\ell}<0,
\qquad \ell\ge2.
\label{eq:supp-negative-hessian}
\end{equation}
Because the round background diagonalizes the quadratic form in scalar harmonics, any nonzero smooth perturbation with support in \(\ell\ge2\) has strictly negative constrained York Hessian.  The local result is therefore generic in the physical nonspherical sector.

Finally, the universal coefficient \(-1/(4G)\) arises only at the smooth Euclidean bolt.  For a regulated tube \(\rho=\epsilon\) around a non-extremal horizon,
\(K\sim1/\epsilon\), \(d\Sigma\sim\kappa\epsilon\,d\tau dA_H\), and regularity gives \(\beta\kappa=2\pi\).  Thus
\begin{equation}
\lim_{\epsilon\to0}I_{\rm GHY}^{(H)}=-\frac{A_H}{4G},
\end{equation}
which is distinct from the geometry-dependent York-boundary coefficient \(C_Y\).

\section{Fixed-charge Einstein--Maxwell extension}
\label{supp:RN}
The geometric focusing argument and the local entropy-Hessian calculation extend to the static Einstein--Maxwell sector at fixed electric charge.  We treat the two ingredients separately.  The focusing step depends on the spacetime Ricci tensor, while the local stability step is obtained from the direct constrained second variation of the York Gibbons--Hawking--York boundary term.

The bulk equations are
\begin{equation}
R_{ab}-\frac{1}{2}R g_{ab}+\Lambda g_{ab}=8\pi G\,T^{\rm EM}_{ab},
\label{eq:einstein-maxwell}
\end{equation}
with
\begin{equation}
T^{\rm EM}_{ab}=F_{ac}F_b{}^{c}-\frac{1}{4}g_{ab}F_{cd}F^{cd}.
\label{eq:maxwell-stress}
\end{equation}
Equivalently,
\begin{equation}
R_{ab}=\frac{2\Lambda}{D-2}g_{ab}
+8\pi G\left(T^{\rm EM}_{ab}-\frac{1}{D-2}T^{\rm EM}g_{ab}\right).
\label{eq:ricci-einstein-maxwell}
\end{equation}
For a static electrically charged black hole the Maxwell field is radial.  In an orthonormal frame adapted to the static geometry,
\begin{equation}
F_{\hat t\hat r}=E(r),
\end{equation}
and
\begin{equation}
T^{\rm EM}_{\hat t\hat t}=\frac{1}{2}E^2,
\qquad
T^{\rm EM}_{\hat r\hat r}=-\frac{1}{2}E^2,
\qquad
T^{\rm EM}=\frac{D-4}{2}E^2.
\end{equation}
Let $u^a$ be the timelike unit vector entering the sheet-focusing argument.  Equation~\eqref{eq:ricci-einstein-maxwell} gives
\begin{align}
R_{ab}u^a u^b
&=-\frac{2\Lambda}{D-2}
+8\pi G\left(T^{\rm EM}_{ab}u^a u^b+\frac{1}{D-2}T^{\rm EM}\right)\notag\\
&=-\frac{2\Lambda}{D-2}
+8\pi G\frac{D-3}{D-2}E^2>0,
\label{eq:charged-timelike-ricci}
\end{align}
since $\Lambda<0$.  Hence the Ricci contribution $-R_{ab}u^a u^b$ in the timelike Raychaudhuri equation has the focusing sign.  On the admissible contracting geodesic branch,
\begin{equation}
\Theta_{(u)}=(D-2)\frac{d\phi}{d\tau},
\qquad
\Theta_{(u)}<0\quad\Longrightarrow\quad\phi<0,
\end{equation}
so the proper negative conformal direction required by the rigidity argument persists at fixed charge.

For the local York variation consider the $D$-dimensional Reissner--Nordstr\"om--AdS geometry
\begin{equation}
f(r)=1-\frac{\mu}{r^{D-3}}+\frac{q^2}{r^{2D-6}}+\frac{r^2}{l^2},
\label{eq:rnads-f}
\end{equation}
and a round Euclidean York boundary $\mathcal B_Y=S^1_\beta\times S_Y$ at $r=R>r_h$.  Put
\begin{equation}
d\equiv D-2,\,\,
N_Y=\sqrt{f(R)},
\,\,
a=\frac{f'(R)}{2\sqrt{f(R)}},
\,\,
b=\frac{\sqrt{f(R)}}{R},
\end{equation}
where $a$ is the Euclidean-time principal curvature and $b$ is each angular principal curvature.  The exact GHY first variation derived in Sec.~\ref{supp:york} remains
\begin{equation}
\delta I_{Y}^{(Q)}
=-\frac{\beta}{8\pi G}\int_{S_Y}N_YQ_Y^{(Q)}\,\phi\,dA_Y,
\label{eq:charged-exact-york-first}
\end{equation}
with
\begin{equation}
Q_Y^{(Q)}\equiv K^2-K_{ij}K^{ij}-R^{(D)}_{nn}.
\end{equation}
Direct substitution of Eq.~\eqref{eq:rnads-f} gives
\begin{equation}
Q_Y^{(Q)}=
\frac{d(d-1)}{R^2}
+\frac{(d+1)^2}{l^2}
-(d-1)\frac{q^2}{R^{2d}}.
\label{eq:charged-QY}
\end{equation}
At the round reference surface $N_Y$, $Q_Y^{(Q)}$, and the spatial mean curvature $H_Y=d\sqrt{f(R)}/R$ are constant.  The first-order proportionality
\begin{equation}
\delta I_Y^{(Q)}=C_Y^{(Q)}\,\delta A_Y,
\qquad
C_Y^{(Q)}=-\frac{\beta N_Y}{8\pi G}\frac{Q_Y^{(Q)}}{H_Y},
\label{eq:charged-first-order-coefficient}
\end{equation}
is therefore used only at the round surface to fix the constrained stationary point.

Let us again introduce the constrained York functional
\begin{equation}
\mathcal I_Y^{(Q)}
=I_Y^{(Q)}+\eta_Q(V_Y-V_0),
\label{eq:charged-constrained-york}
\end{equation}
with $\delta V_Y=\int_{S_Y}\phi\,dA_Y$.  Stationarity fixes
\begin{equation}
\eta_Q=\frac{\beta}{8\pi G}\bigl(N_YQ_Y^{(Q)}\bigr)_0
=-C_Y^{(Q,0)}H_Y.
\label{eq:charged-eta}
\end{equation}
At this constrained critical point the Hessian is independent of the second-order extension of the one-parameter deformation.  Transporting the first-order normal speed with $\delta\phi=0$ and differentiating Eq.~\eqref{eq:charged-exact-york-first} together with the volume term gives
\begin{equation}
\delta^2\mathcal I_Y^{(Q)}
=-\frac{\beta}{8\pi G}\int_{S_Y}
\phi\,\delta\!\left(N_YQ_Y^{(Q)}\right)dA_Y.
\label{eq:charged-direct-hessian}
\end{equation}
All measure and acceleration terms multiply the vanishing first-variation coefficient fixed by Eq.~\eqref{eq:charged-eta}.  Thus Eq.~\eqref{eq:charged-direct-hessian} is the direct constrained second variation.

The linearization has the same derivative structure as in Sec.~\ref{supp:york}.  In the RN--AdS background $R^{(D)}_{nA}=0$, so the variation of the normal under an angular shape deformation produces no additional first-derivative term in $\delta R^{(D)}_{nn}$.  The remaining matter contribution is zeroth order and is contained in the radial derivative of $Q_Y^{(Q)}$.  Hence
\begin{equation}
\delta Q_Y^{(Q)}
=-2\bigl[a+(d-1)b\bigr]\Delta_{S_Y}\phi
+\bigl(n\!\cdot\!\nabla Q_Y^{(Q)}\bigr)\phi,
\label{eq:charged-delta-Q}
\end{equation}
and therefore
\begin{equation}
\delta\!\left(N_YQ_Y^{(Q)}\right)
=-2N_Y\bigl[a+(d-1)b\bigr]\Delta_{S_Y}\phi
+\bigl(n\!\cdot\!\nabla(N_YQ_Y^{(Q)})\bigr)\phi.
\label{eq:charged-delta-NQ}
\end{equation}
For a scalar harmonic
\begin{equation}
-\Delta_{S_Y}\phi_\ell=\frac{\lambda_\ell}{R^2}\phi_\ell,
\qquad
\lambda_\ell=\ell(\ell+d-1),
\end{equation}
Eq.~\eqref{eq:charged-direct-hessian} becomes
\begin{equation}
\delta^2\mathcal I_{Y,\ell}^{(Q)}
=-\frac{\beta}{8\pi G}\,
\mathcal K_\ell^{(Q)}(R)\,\|\phi_\ell\|^2,
\label{eq:charged-mode-hessian}
\end{equation}
where
\begin{equation}
\begin{split}
\mathcal K_\ell^{(Q)}
={}&\frac{\lambda_\ell}{R^2}
\left[f'(R)+\frac{2(d-1)}{R}f(R)\right]
+\frac{f'(R)}{2}Q_Y^{(Q)}\\&{}
+f(R)\bigl(Q_Y^{(Q)}\bigr)'.
\end{split}
\label{eq:charged-Kell}
\end{equation}

The sign of Eq.~\eqref{eq:charged-Kell} can be shown directly throughout the nonextremal exterior. Let us define
\begin{equation}
\begin{split}
&w\equiv\frac{q^2}{R^{2d-2}},
\qquad
t\equiv\frac{R^2}{l^2},\\&
A\equiv R^2Q_Y^{(Q)}
=d(d-1)+(d+1)^2t-(d-1)w.
\end{split}
\end{equation}
The horizon condition together with nonextremality gives
\begin{equation}
\frac{q^2}{r_h^{2d-2}}
\leq 1+\frac{d+1}{d-1}\frac{r_h^2}{l^2}.
\label{eq:rnads-charge-bound}
\end{equation}
For $R>r_h$ this implies
\begin{equation}
w<1+\frac{d+1}{d-1}t,
\end{equation}
and therefore
\begin{equation}
A>(d-1)^2+d(d+1)t>0.
\label{eq:charged-A-positive}
\end{equation}
Moreover, for the outer nonextremal horizon $f'(r_h)>0$.  Since
\begin{equation}
\frac{d}{dr}\left[r^{2d-1}f'(r)\right]
=(d-1)^2\mu r^{d-2}+\frac{4d}{l^2}r^{2d-1}>0,
\end{equation}
it follows that $f'(R)>0$ for every $R>r_h$.

Using
\begin{equation}
R^3\bigl(Q_Y^{(Q)}\bigr)'=2d(d-1)(w-1),
\end{equation}
Eq.~\eqref{eq:charged-Kell} may be rearranged as
\begin{equation}
R^3\mathcal K_\ell^{(Q)}
=Rf'(R)\left(\lambda_\ell+\frac{A}{2}\right)
+2(d-1)f(R)\left[\lambda_\ell+d(w-1)\right].
\label{eq:charged-Kell-positive-form}
\end{equation}
Every factor on the right-hand side is positive in the nonextremal exterior.  Indeed, $f(R)>0$, $f'(R)>0$, and $A>0$, while for $\ell\ge2$ and $w\ge0$,
\begin{equation}
\lambda_\ell+d(w-1)
\geq \lambda_2-d=d+2>0.
\end{equation}
Consequently,
\begin{equation}
\mathcal K_\ell^{(Q)}(R)>0,
\qquad R>r_h,
\qquad \ell\ge2,
\end{equation}
and hence
\begin{equation}
\delta^2\mathcal I_{Y,\ell}^{(Q)}<0,
\qquad \ell\ge2.
\label{eq:charged-outer-hessian-negative}
\end{equation}
By harmonic decomposition the same strict negativity holds for every nonzero smooth perturbation supported in the physical nonspherical sector.

The fixed-charge electromagnetic variational principle is implemented by the Maxwell action
\begin{equation}
I_{\rm M}=-\frac{1}{16\pi G}\int_{\mathcal M}F_{\mu\nu}F^{\mu\nu}\sqrt{-g}\,d^Dx
\end{equation}
and the standard electric Legendre boundary term
\begin{equation}
I_{\rm bdy}^{(Q)}=-\frac{1}{4\pi G}
\int_{\partial\mathcal M}\sqrt{h}\,n_\mu F^{\mu\nu}A_\nu.
\end{equation}
Its variation fixes the normal electric flux, so there is no independent electromagnetic work term along an admissible fixed-$Q$ shape variation.  On the fixed-$Q$ Brown--York microcanonical saddle,
\begin{equation}
S_{\rm RN}
=iI_{\rm mc}^{(Q),\,\rm saddle}
=\frac{\mathcal A_H}{4G}.
\end{equation}
At the round RN--AdS reference point the first variation of the scalar thermodynamic volume is rotationally invariant and therefore lies entirely in the monopole sector.  A pure $\ell\ge2$ tangent consequently satisfies $\delta V_{\rm th}=0$ at first order, while the required second-order monopole correction keeps the saddle family on $V_{\rm th}=V_*$.  The York volume in Eq.~\eqref{eq:charged-constrained-york} is only the local constraint coordinate at the round reference point.  Therefore the nonspherical constrained Hessian of the fixed-$(V_{\rm th},Q)$ microcanonical saddle is the direct York shape Hessian above,
\begin{equation}
\left.S_{\rm RN}''(0)\right|_{V_{\rm th},Q}
=\left.\frac{d^2\Gamma_{\rm mc}^{(Q)}}{ds^2}\right|_{0,V_{\rm th},Q}
=\delta^2\mathcal I_{Y,\ell}^{(Q)}<0,
\qquad \ell\ge2.
\label{eq:charged-entropy-hessian}
\end{equation}
Thus the round RN--AdS horizon is a strict local entropy maximum in the admissible nonspherical sector at fixed thermodynamic volume and fixed electric charge.

Finally, the thermodynamic volume of the static charged family is
\begin{equation}
\mathcal V_{\rm th}
=\left(\frac{\partial M}{\partial P}\right)_{S,Q}
=\frac{\mathcal A_{D-2}r_h^{D-1}}{D-1}.
\label{eq:rnads-volume}
\end{equation}
Since $S\propto r_h^{D-2}$, fixing $S$ fixes $r_h$, and the charge-dependent term in the mass does not contribute to $(\partial M/\partial P)_{S,Q}$.  The static fixed-$Q$ thermodynamic-volume constraint is therefore represented by the same areal volume that labels the spherical RN--AdS family.

The charged extension consequently preserves both ingredients of the local argument: Einstein--Maxwell focusing supplies the required negative conformal direction, and the direct constrained York Hessian is strictly negative for every nonspherical mode $\ell\ge2$ throughout the non-extremal exterior.

\bibliography{bib}
\bibliographystyle{utphys1}

\end{document}